\documentclass{sigchi}

\toappear{}

\usepackage{balance}  
\usepackage{graphics} 
\usepackage{times}    
\usepackage{url}
\usepackage[activate]{microtype}
\usepackage{nameref}
\makeatletter
\def\url@leostyle{%
  \@ifundefined{selectfont}{\def\UrlFont{\sf}}{\def\UrlFont{\small\bf\ttfamily}}}
\makeatother
\def\pprw{8.5in}
\def\pprh{11in}

\usepackage[pdftex]{hyperref}
\hypersetup{
pdftitle={SIGCHI Conference Proceedings Format},
pdfauthor={LaTeX},
pdfkeywords={SIGCHI, proceedings, archival format},
bookmarksnumbered,
pdfstartview={FitH},
colorlinks,
citecolor=black,
filecolor=black,
linkcolor=black,
urlcolor=black,
breaklinks=true,
}

\begin{document}

\title{The state of play of ASC-Inclusion: An Integrated\\ Internet-Based Environment for Social Inclusion of\\ Children with Autism Spectrum Conditions}

\numberofauthors{20}
\author{
 \alignauthor Bj{\"o}rn Schuller,\\
Erik Marchi\\
       \affaddr{Machine Intelligence and Signal Processing group}\\
       \affaddr{Technische Universit{\"a}t M{\"u}nchen}\\
       \affaddr{80333 M{\"u}nchen, Germany}\\
       \email{schuller@tum.de}
       \email{erik.marchi@tum.de}
\alignauthor Simon Baron-Cohen,\\
Helen O'Reilly\\
Delia Pigat\\
      \affaddr{Autism Research Centre}\\
      \affaddr{University of Cambridge}\\
      \affaddr{Cambridge, UK}\\
      \email{sb205@cam.ac.uk}
      \email{heo24@cam.ac.uk}
			\email{dp467@medschl.cam.ac.uk}
\alignauthor Peter Robinson,\\
Ian Davies\\
      \affaddr{Computer Laboratory}\\
      \affaddr{University of Cambridge}\\
      \affaddr{Cambridge, UK}\\
      \email{peter.robinson@cl.cam.ac.uk}
      \email{Ian.Davies@cl.cam.ac.uk}
}


\maketitle

\begin{abstract}
Individuals with Autism Spectrum Conditions (ASC) have marked difficulties using verbal and non-verbal communication for social interaction. 
The running ASC-Inclusion project aims to help children with ASC by allowing them to learn how emotions can be expressed and recognised via playing games in a virtual world. The platform includes analysis of users' gestures, facial, and vocal expressions using standard microphone and web-cam or a depth sensor, training through games, text communication with peers, animation, video and audio clips. We present the state of play in realising such a serious game platform and provide results for the different modalities.
\end{abstract}



\section{Introduction}
Three decades of research have shown that children and adults with Autism Spectrum Conditions (ASC) experience significant difficulties recognising and expressing emotions and mental states \cite{baron1997mindblindness}.
These difficulties are apparent when individuals with ASC attempt to recognise emotions from facial expressions \cite{celani1999understanding, deruelle2004spatial, golan2006cambridge, hobson1986autistic}, from vocal intonation \cite{boucher2003voice, golan2007reading}, from gestures and body language \cite{grezes2009failure, philip2010deficits}, and from the integration of multi-modal emotional information in context \cite{golan2008reading, silverman2010speech} 
Limited emotional expressiveness in non-verbal communication is also characteristic in ASC, and studies have demonstrated individuals with ASC have difficulties directing appropriate facial expressions to others \cite{kasari1990affective, kasari1993affective, kasari2001social}, modulating their vocal intonation appropriately when expressing emotion \cite{macdonald2006recognition, mccann2003prosody, paul2005perception} and using appropriate gestures and body language \cite{attwood1998asperger}. 
Integration of these non-verbal communicative cues with speech is also hampered \cite{de2010conversational}. 
Individuals with ASC lack the sense of social reciprocity and fail to develop and maintain age appropriate peer relationships \cite{bell1994dsm}, \cite{world1993icd}. Current findings suggest 1\,\% of the population might fit an ASC diagnosis \cite{baron2009prevalence}. 

The use of Information Communication Technology (ICT) with individuals with ASC has flourished in the last decade for several reasons: the computerised environment is predictable, consistent, and free from social demands, which individuals with ASC may find stressful. Users can work at their own pace and level of understanding, and lessons can be repeated over and over again, until mastery is achieved. In addition, interest and motivation can be maintained through different and individually selected computerised rewards \cite{bishop2003,moore2000computer,parsons2002}. 
For these reasons, and following the affinity individuals with ASC show for the computerised environment, dozens of ICT programs, teaching various skills to this population were created. However, most of these tended to be rather specific (e.\,g., focusing only on recognition of facial expressions from still photos) and low budget, and have not been scientifically evaluated \cite{golan2007assistive}. 
ICT programs that teach socio-emotional communication have been evaluated including \textit{I can Problem-Solve}; others aim to teach social problem solving \cite{bernardopitz2001}, such as  
\textit{FEFFA}, e.\,g., by emotion recognition from still pictures of facial expressions and strips of the eye region \cite{bolte2006facial}. \textit{Emotion Trainer} teaches emotion recognition of four emotions from facial expressions \cite{silver2001evaluation}; \textit{Let’s Face It} is teaching emotion and identity recognition from facial expressions \cite{tanaka2010}, and \textit{Junior Detective} program combines ICT with group training in order to teach social skills to children with ASC \cite{Beaumont2008}. These examples demonstrate how focused most ICT solutions are in their training, focusing mostly on emotion recognition from facial expressions and contextual situations, with very little attention given to emotional gestures of emotional voices. In addition, there were no reports of ICT programs teaching emotional expressiveness.
Further training programs such as \textit{Mind Reading} implement an interactive guide to emotions and teaches recognition of 412 emotions and mental states, systematically grouped into 24 emotion groups, and 6 developmental levels (from the age of four years to adulthood). The evaluations of Mind Reading resulted in limited generalization when adults with ASC used the software \cite{CambridgeJournals:420155,golan2007assistive}. However, when 8-11 year old children with ASC had used it, improved generalization was found.

In the last decade, with the rapid development of internet-based communication, web applications have been increasingly used for social interaction, forming online communities and social networks. Anecdotal reports of the emergence of online ‘autistic communities’, and the use of forums, chat rooms and virtual-worlds, show the great promise the internet holds for better inclusion and social skills training for adolescents and adults with ASC \cite{brownlow2006constructing,jordan2010evolution}. However, there has been no scientifically documented attempt to use the internet for structured training of socio-emotional communication for individuals with ASC. Furthermore, since intervention into ASC has been shown to be more effective when provided early in life, using the internet as a platform for the support of children with ASC could significantly promote their social inclusion.

Virtual Environments (VE) form another domain with immense possibilities for those with ASC and related social difficulties. VE are artificial computer generated three-dimensional simulations and come in single- or multi-user forms. In either format, the user can operate in realistic scenarios to practice social skills, conversations, and social problem solving. Moore and colleagues investigated VE and children and youth with ASC and found that over 90\,\% of their participants used the VE to recognise basic emotions \cite{moore2005collaborative}. Other studies have also shown the potential for those with ASC to use VE for socio-emotional skills training and for other learning purposes \cite{parsons2000development,parsons2002}.

The ASC-Inclusion project\footnote{http://www.asc-inclusion.eu} \cite{Schuller13-AIE} -- dealt with herein -- suggests advanced ICT-enabled solutions and serious games for the
empowerment of children with ASC who are at high risk of social exclusion.
It combines several state-of-the-art technologies in one
comprehensive game environment, including analysis of users' gestures, facial, and vocal expressions,
training through games, text chatting, animation, video and audio clips. 
Despite the innovative technologies involved, the ASC-Inclusion is aimed for home use. 
Though designed to assist children with ASC, ASC-Inclusion could serve other population groups
characterised by deficient emotional understanding and social skills, such as children with learning
difficulties \cite{bauminger2005social}, attention deficit and hyperactivity disorder (ADHD) \cite{da2009emotion}, behavioural and conduct
problems \cite{stevens2001recognition}, or socio-emotional difficulties \cite{pollak2000recognizing}.

The remainder of this paper is structured as follows: first, a detailed description of the user requirements and specification is given (Section \nameref{sec:requirements}); then we describe the three modalities namely face, voice and body gesture (Sections \nameref{sec:face}, \nameref{sec:voice}, \nameref{sec:gestures}) and the platform (Section \nameref{sec:platform}). We next comment on content creation (Section \nameref{sec:content}) before concluding the paper in Section \nameref{sec:conclusions}.

\section{User Requirements}
\label{sec:requirements}
Previous literature reports social communication difficulties in individuals with ASC as well as enhanced abilities in other non-social areas such as systemising \cite{baron2005empathizing}. This provided us with a greater insight into the social, cognitive and behavioural abilities of these potential end users. Review of the design, content and effectiveness of the Mind Reading DVD as an intervention \cite{CambridgeJournals:420155} and The Transporters DVD \cite{golan2010enhancing} which have shown to enhance the ability of children with ASC to recognise emotions, permitted identification of a method through which learning material can be presented in an engaging manner which capitalises on the enhanced systemising skills seen in individuals with ASC.

A preliminary set of focus groups in Israel and Sweden were observed to help identify the preferences and interaction style of children with ASC when using virtual environments. The results from Hebrew user focus groups found that the children enjoyed using the virtual environments and the majority of children could navigate independently in the environment, activate different functions of the VE and change the settings.  The results from these focus groups found that the games that were easy to understand and had a good game concept were rated highest by the children. It was noted that a number of the games presented to the children and which the children enjoyed playing could be easily adapted to become emotion recognition games, but that better reward systems and the ability to change difficulty level was necessary to keep the children engaged and motivated.

Further focus groups were observed to asses if the children could maintain attention to a teacher talking about emotions and understand how emotional dimensions are expressed in the voice. The results from this study showed that the maturity level appears to affect children’s ability to maintain attention and comprehend the topic of emotions in voices. Additional instructions are required for younger children when teaching emotions in the voice. Illustrative examples may help the children to grasp the concept of the different vocal dimensions such as pitch and vocal speed.

A second batch of focus groups were conducted in Israel with the main aims of observing the children's use and reactions to the virtual environment, the four educational games, and the `anger' lessons provided in the current ASC-Inclusion platform by assessing the level of attention and learning. Additionally, the first version of the voice analyser as further described below was tested. Generally, the Focus Group children loved the virtual environment. They responded in a positive way to the environment and navigated freely and instinctively through it. The children in particular liked the functions of the environment that they could design themselves such as the avatar character and their virtual home, these features of the environment are significantly motivational. The children also enjoyed being able to equip their avatar character. All the children listened throughout the ‘anger’ lessons, and the material was understood. Most of the children also liked the emotional games with particular preference to ``Emotion Tetris'' and ``Emotion Memory Game''. All the children liked 
the activity with the voice analyser software. They related to the plots and understood how to modulate vocal parameters after a short explanation from the experimenter. The children had fun playing with their voices and influencing the plots shown to them based on their speech acoustics.

Further user testing was conducted with additional games embedded into the platform such as ``Who wants to be a Millionaire'' and ``Everyday Stories''. The former assesses the child on the level of learning in the unit just completed; performance on this quiz determines the child's progression to the next learning unit.
The latter allows the child to practice and demonstrate the knowledge he/she acquired in a learning unit. The child deals with integrative presentation of the social situations, including face, body, and voice materials. The results show that in the former game, the percentage of questions correctly answered increased with age. The information gathered from review of literature  and technologies developed for children with ASC, as well as the results from the focus groups have been fundamental in designing both the visual and functional aspects of the VE to match the needs of the intended user -- children with ASC 5--10 years of age.

\section{Face Analysis}
\label{sec:face}
The previous Facial Affect Computation Engine (FACE) \cite{Schuller13-AIE} is using a dimensional model of affect to fit with the other inference systems from voice and from body motion. The dimensional model requires a different approach to affective inference, building on our earlier experience with categorical inference \cite{kaliouby2005real}. We want to model the affect in a continuous space rather than as discrete categories as in our initial prototype.  Furthermore, we want to model the temporal relationships between each time step, since emotion has temporal properties and is not instantaneous.  The constrained local model (CLM) tracker \cite{3d_6247980} has been retained, but the mental state inference now applies linear support vector regression (SVR) for inference.
The dimensional inference system uses the CLM tracker to extract 34 features from live video of the face.  These include per-frame descriptors of the facial expression and measures of variation in head pose over time.  SVR was trained on the AVEC 2012 corpus \cite{Schuller12-A2T} to predict values for valence and arousal.  
The output from the FACE engine is encoded in EmotionML \cite{schroeder2010emotion} and delivered through an ActiveMQ \footnote{“Apache ActiveMQ,” http://activemq.apache.org/.} communication infrastructure, allowing integration with vocal analysis and gesture analysis to provide multi-modal inference in the central platform.

\begin{figure}[t]
\centering
\begin{tabular}{lr}
\includegraphics[width=8cm]{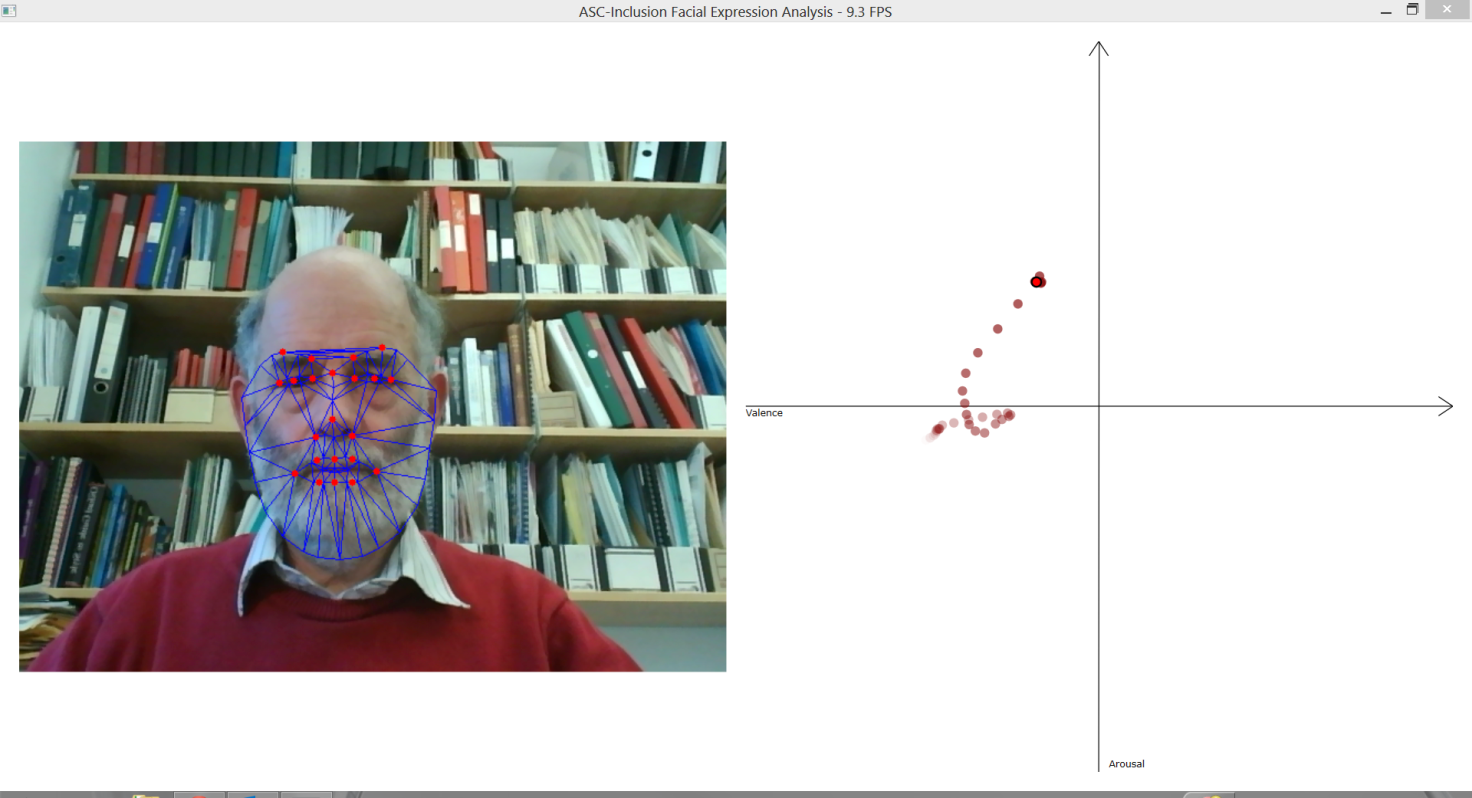}
\end{tabular}
\caption{Face analyser.}
\label{fig:face}
\end{figure}

\section{Voice Analysis}
\label{sec:voice}
\begin{figure}[t]
\centering
\begin{tabular}{lr}
\includegraphics[width=8cm]{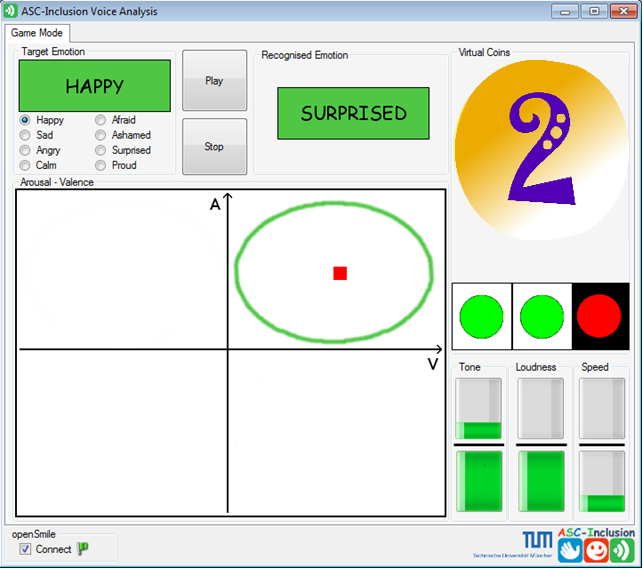}
\end{tabular}
\caption{Voice analyser.}
\label{fig:voice}
\end{figure}

For the training of a voice analyser, a data set of prototypical emotional utterances containing sentences spoken in English, Hebrew, and Swedish by a total number of 60 children has been collected with typically developing children and children with ASC. A complete description of the database containing sentences in Hebrew is given in \cite{Marchi12-EIT}.
Considering the outcome of the evaluations described in \cite{Marchi12-EIT} and \cite{Marchi12-SEA}, we defined potential features and descriptors that are relevant for emotion perception, such as energy by the sum of auditory spectrum at different frequency bands (from 20\,Hz to 8\,kHz) and root-mean-square signal frame energy, pitch by fundamental frequency contour, and duration by modelling temporal aspects of F0 values, such as the F0 onset segment length.
Applying the openSMILE audio feature extractor \cite{Eyben10-OTM,Eyben13-RDI}, features are extracted and tracked over time:  
%
%
in order to assess a child's performance in expressing emotions via speech, the extracted parameters are compared to the respective parameters extracted from pre-recorded prototypical utterances.
The vocal expression evaluation system is shown in Figure \ref{fig:voice}. 
In the ``Target emotion'' box the player chooses the emotion she or he wants to play with. Once the emotion has been selected, a reference emotion expression is played back to the child and the according quadrant of the arousal / valence plane  is highlighted. 
Then, the child is prompted to repeat the selected emotion. According to the expressed emotion the evaluation system is providing a visual feedback on the aurousal / valence plane via a red dot indicating the coordinates of the expressed emotion in the two dimensional space. Additionally, in the ``Recognised Emotion Group'' box the detected emotion is expressed.
According to the correctness of the expressed emotion virtual coins are earned and on the bottom right part of the GUI a corrective feedback is shown. The `traffic light' on the top of the gauge bars indicates if the extracted parameters are distant or close to the reference values. If a green light is shown the child’s expression is close to the reference; the red light indicates high distance between the reference value and the extracted one.

The output from the voice analyser is also encoded in EmotionML \cite{schroeder2010emotion} and delivered through the ActiveMQ communication infrastructure, allowing integration with the face analysis and the gesture analysis to provide multi-modal inference in the central platform.

\section{Gesture Analysis}
\label{sec:gestures}
Full-body expressive features to be analysed in the game are inspired by several sources: biomechanics, computer vision, experimental psychology, and humanistic theories, including Laban's Effort Theory \cite{laban1971mastery}. The list of candidate features is the following:
kinetic energy of body parts (hands, head, upper body), symmetry of hands movements and posture with respect to the vertical axis and forward/backward, forward/backward leaning of upper part of the body, directness (direct/flexible in terms of Laban's Effort), impulsivity, fluidity, light/heavy gesture quality, the relative position between head and shoulder, openness of the posture (contraction/expansion), rigidity and swinging of the whole body or of body parts.

A dataset of 15 people expressing the six basic emotions (happiness, anger, sadness, disgust, fear, and surprise) was collected. Each person was asked to express one of the six emotions with their body from three to seven times in front of a Microsoft Kinect sensor. On average, each recording session lasted about 10 minutes. After the recording sessions, each video was manually segmented to find the prototypical expressive gestures of each emotion. Those segments were used to evaluate the data with human subjects. 60 judges labelled 10 segments, each. The material was used to train SVM for the classification modules.
A real-time feature extractor has been developed using the EyesWeb \footnote{\url{http://www.infomus.org/eyesweb_ita.php}} XMI development environment; the existing library for gesture analysis has been extended and modified according to the descriptors and feature set introduced before. The EyesWeb development environment main characteristics are: the multi-modal input processing capabilities, modularity, flexibility and extensibility. The key features are as follows:
Multimodal input support, where data from an arbitrary input stream (file, webcams, RGB-D cameras, microphones, sensors, etc.) is pushed through the processing chain sample by sample and frame by frame,  
modularity, which allows modifying the extracted features set, and easy addition of new feature extractors via a dedicated API, and finally easy configuration of feature extractor parameters that can even be tuned during run-time through the EyesWeb XMI development environment.

\begin{figure*}[t]
\centering
\begin{tabular}{lr}
\includegraphics[width=16cm]{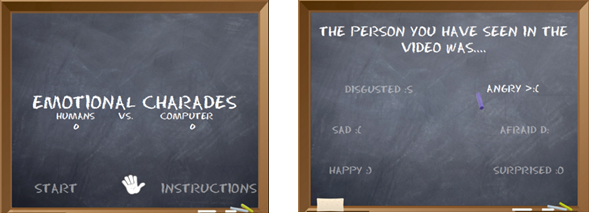}
\end{tabular}
\caption{Body gesture analyser.}
\label{fig:body}
\end{figure*}

Higher level computations such as machine learning techniques are carried by the MetaEyesWeb external module. This adds to the Python programming language the possibility to remotely communicate with the EyesWeb Environment, controls the developed applications execution and tunes directly the module parameters through a dedicated communication API. MetaEyesWeb Mobile enables the use of Python scripts and thus the integration of related machine learning libraries.
Output from the body analyser is again encoded in EmotionML \cite{schroeder2010emotion} and delivered through the ActiveMQ communication infrastructure, allowing integration with face analysis and voice analysis to provide multi-modal inference in the central platform.
The framework described so far was used as the main part of two game demonstrations. Both the games perform a live automatic emotion recognition, and interact with the user by asking her to guess an emotion and to express an emotion with her body (Figure \ref{fig:body}).

\section{Platform}
\label{sec:platform}
\begin{figure}[t]
\centering
\begin{tabular}{lr}
\includegraphics[width=8cm]{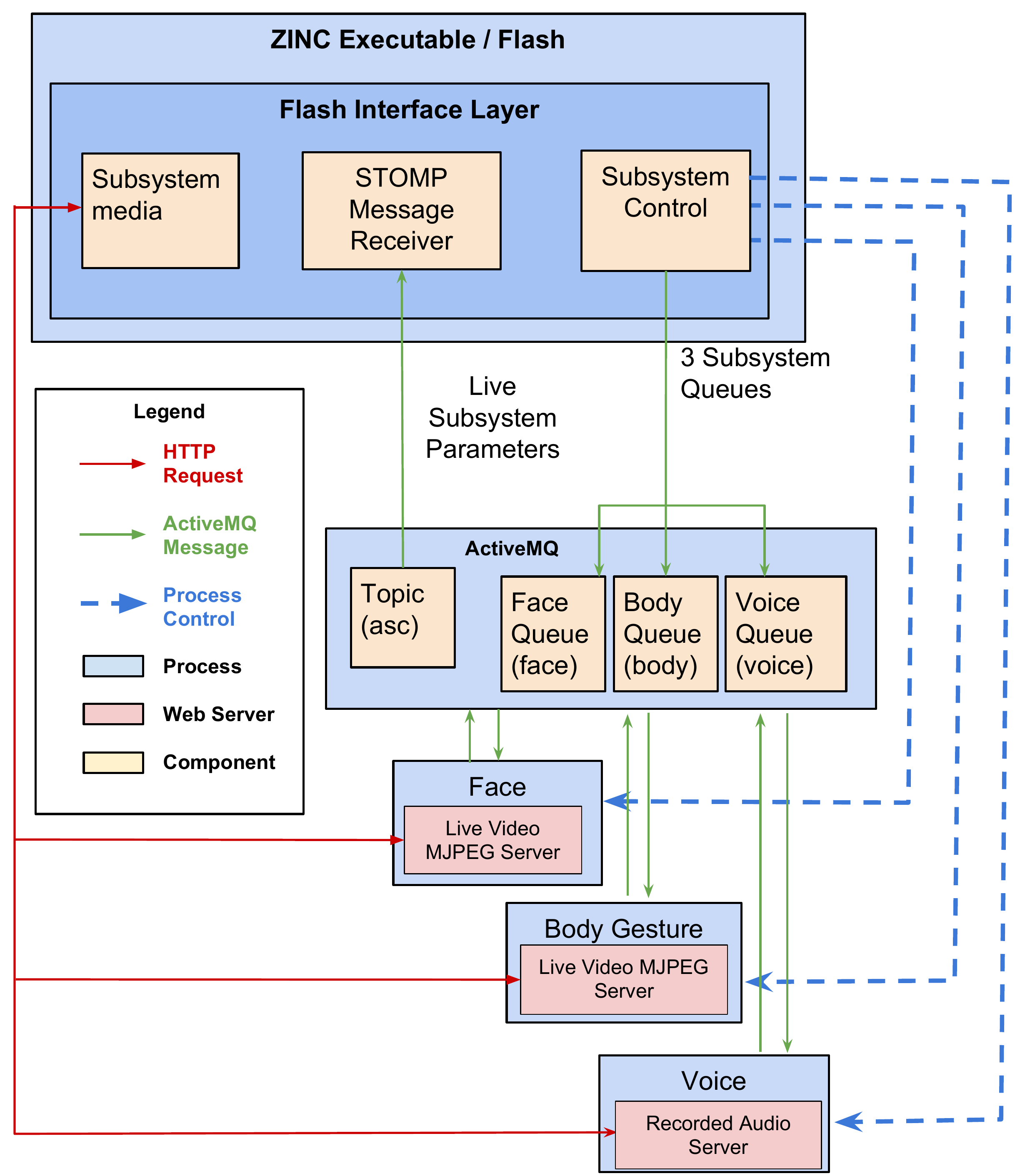}
\end{tabular}
\caption{Integrated platform architecture.}
\label{fig:platform_integrated}
\end{figure}

The general architecture presented in \cite{Schuller13-AIE} has been updated and extended. An overview of the new integrated platform architecture is given in Figure \ref{fig:platform_integrated}. 
The container of the  game engine in flash is ZINC\footnote{http://www.multidmedia.com/software/zinc/}; the container loads the game environment and the subsystem’s services. The flash game engine is lying under the game logic and the user interface. It is responsible for the communication with the different services through Apache ActiveMQ by applying Streaming Text Oriented Messaging Protocol\footnote{http://stomp.github.io} (STOMP). 
The communication with the different services are based on three components. The subsystem control component sends control messages to the different services -- the control messages are sent to a unique ActiveMQ queue for each of the services (face, body, voice). When a message is sent to the activeMQ queue, the subsystem service reads it and processes the command. The STOMP message receiver gets XML text messages from ActiveMQ topic based on STOMP protocol. The component connects to ActiveMQ through STOMP and starts receiving messages from the server. Messages are then used by the game engine in order to provide a certain feedback to the user. As above, the XML messages are encoded using the EmotionML standard.
Since the game needs to show the web camera input, or to play back the voice recorded during a game task, the subsystem `media component' obtains such data from the subsystem services. Each service runs a web server which provides (by request) a stream of information from the camera or the recorded audio. The ActiveMQ process is used as the pipe to push the information from the different subsystems to the game engine. ActiveMQ includes one Topic (named “asc”) which contains all classification results and parameters from the different modalities (based on the running services). ActiveMQ also includes three queues (one per subsystem) which contain control messages produced by the platform (start, stop, shutdown, etc.). The three subsystems provide output in the same dimensional space. The subsystems provide the actual values of the recognised emotion in the arousal and valence space and the platform compares them according to the target space.

\begin{figure}[t]
\centering
\begin{tabular}{lr}
\includegraphics[width=7.5cm]{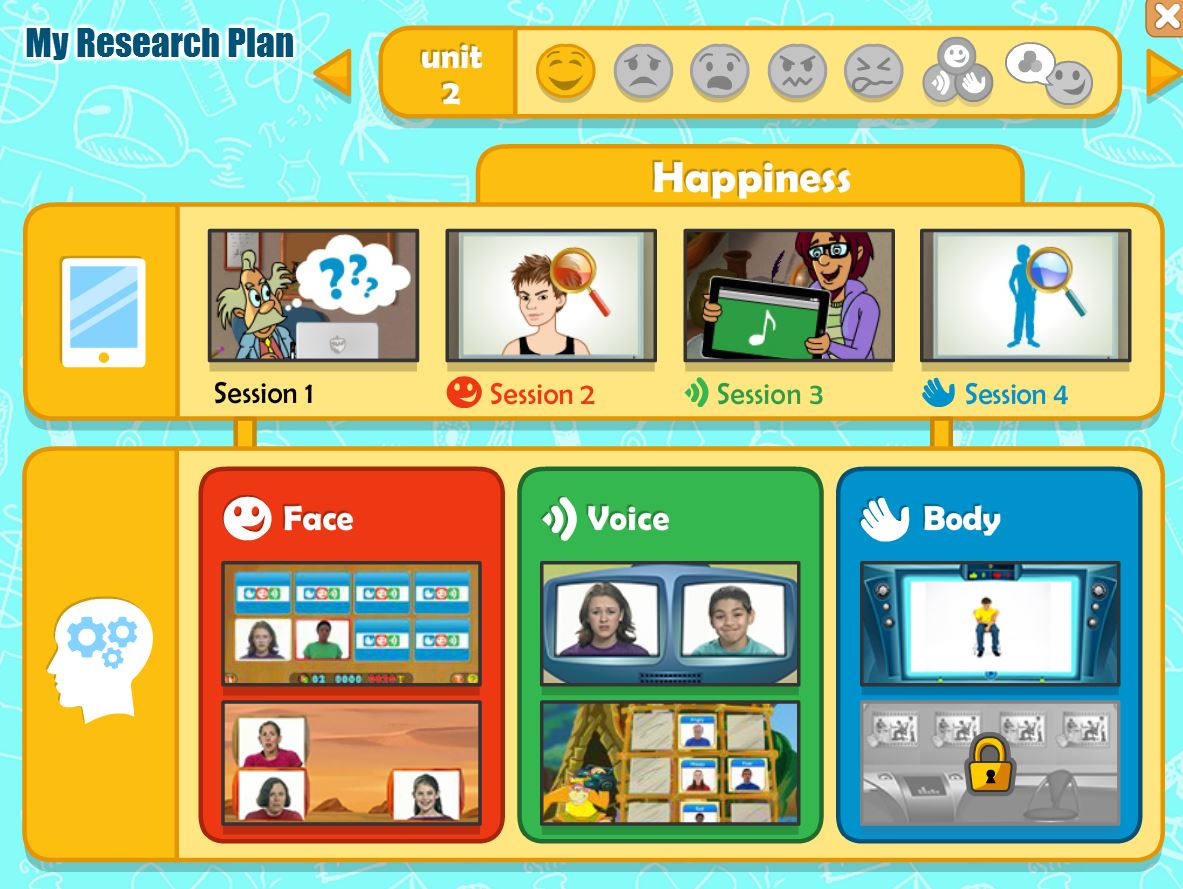}
\end{tabular}
\caption{Learning program manager application user interface.}
\label{fig:learning-manager}
\end{figure}

Besides the platform integration, a new Learning Manager Application (LMA) was developed (cf.\ Figure \ref{fig:learning-manager}). This application is the main tool to control, personalise, and present the learning material to the user. It also enables the navigation within the program including access to the practice games. The LMA enables to track the user's play patterns for data collection and for system improvement purposes. The LMA also manages the rewards granted to the child while going through the program. The `monetary system' is the basis for the VE `economy', planned to motivate the child and encourage long-term engagement. The learning material overall consists of an intro (five animated general introduction sessions, interactive activities), emotion lessons (20 interactive animated `research sessions'), curricular basic games (six simple task games, two for each modality (face, voice, body)), integrative stories (six interactive stories with accompanying activities, two for each of the emotions learning units), advanced games (three cross-modal games, one integrative quiz game), a comprehensive testing game (one per emotion unit, 300 questions arranged in nine levels of difficulty, systematically covering all the unit content), and three semi-curricular dynamic fun games with emotions content.

Advancing in the program by actively participating in `research sessions' (lessons) and playing the practise games to a pre-defined achievement, is the only way to earn virtual money in the VE, and virtual money is needed for virtually anything the child might want to do in the VE apart from `doing research' (learning) -– namely: play non-curricular games, buy goodies, go to fun locations out of the `camp' -- the learning and training activities are embedded in a virtual world themed as a 'research camp', where each user is a 'young researcher' exploring human behaviour -- etc.
This way we hope to maintain an efficient balance between coping with the possibly challenging emotions content, and having the opportunity for recreation and pure fun. This is also meant to prevent a situation of ignoring the program or `getting stuck' on a certain activity or comfort zone.
Additionally, further motivational elements were added, such as the avatar, the camp-home, and the virtual wallet. Four new characters have been designed to serve and support the learning sessions and add interest and fun. Further, four fun locations were designed as out-of-camp `tourist' recreation destinations. The new locations contain fun games with themed content that also has an enrichment element. The child needs to earn the ticket to go there, and usage is limited.

\begin{figure}[t]
\centering
\begin{tabular}{lr}
\includegraphics[width=8cm]{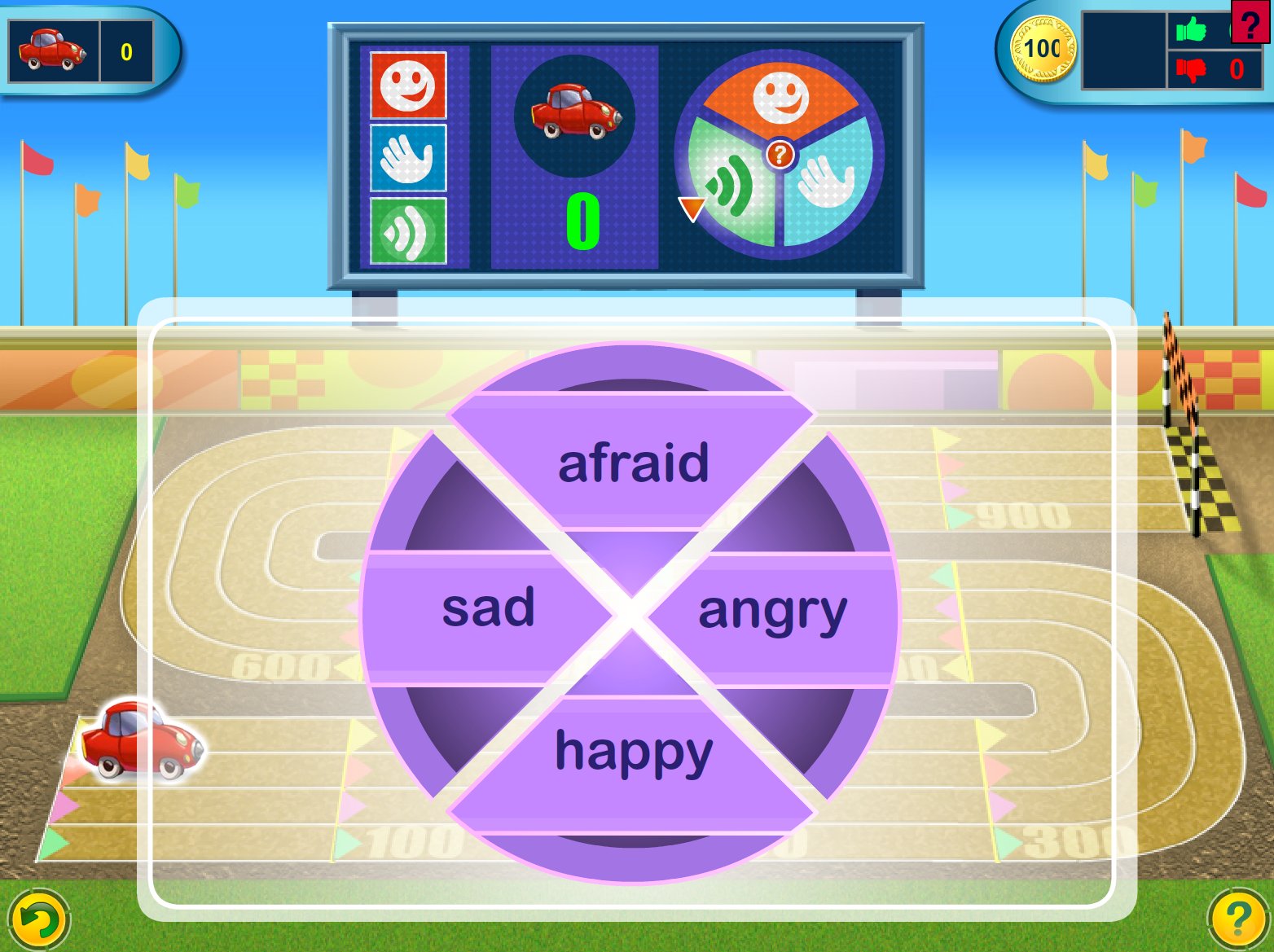}
\end{tabular}
\caption{`Expression' game with face, voice, and body gesture modalities.}
\label{fig:expression-game}
\end{figure}

A first expression game \ref{fig:expression-game} has been developed. It integrates the face, voice, and body language analysis technologies. This game is the main activity in the `expression sub-unit' at the end of each of the learning units. The game is designed as a `race' board game. Each turn the child is asked to express an emotion in a chosen modality. If she/he expressed it well enough, the racing robot advances one step on the board. Whoever gets to the ending point first wins. 

\section{Content Creation}
\label{sec:content}
In order to determine the most important emotions for social interaction in children with ASC, an emotion survey was developed and completed by parents, typically developing adults and parents in particular of children with ASC. 20 emotions -- happy, sad, afraid, angry, disgusted, surprised, excited, interested, bored, worried, disappointed, frustrated, hurt, kind, jealous, unfriendly, joking, sneaky, ashamed, and proud -- were identified by the survey as being the most important for social interaction and selected to be taught through the platform.
Additionally, 496 face stimuli, 82 body gesture stimuli and 95 social scene stimuli previously recorded with professional actors were validated. The stimuli were divided into 21 validation surveys with each survey containing 30-45 emotion stimuli. Each survey was completed by 60 typically developing adults (20 per country), at minimum age of 18 years. For each stimulus validated a percentage of correct recognition score and a chance corrected percentage recognition score was calculated. 
These two scores detail the percentage of participants that correctly identified the emotion. However, the chance corrected score takes into consideration the possibility of the person selecting the correct emotion label by chance. Only emotions with a chance corrected percentage score greater than 50\,\% are included in the platform. A further 20 UK-based voice surveys have been created and validated along side voice recordings in Hebrew and Swedish: As outlined above, to assist with the development of the voice analysis module, a series of emotion phrases were recorded by in UK, Israel, and Sweden. A total of 20 children (10 typically developing and 10 children with ASC) were recorded at each site, in English, Hebrew and Swedish. 
The discrete emotions were portrayed within phrases that the child had to repeat; each phrase was embedded within a story to provide context for the child to express the emotion. Four phrases were written for each of the emotions.

\section{Conclusions}
\label{sec:conclusions}
We introduced the current state of the gaming platform ASC-Inclusion targeted to children aged 5 to 10 years with ASC. User requirements and specification have been refined and further analysed, tailoring the needs of the target audience and creating user scenarios for the platform subsystems. 
The integration of the three subsystem into the game platform was implemented. The technical partners collaborated to design communication protocols and the platform architecture for the main game engine. The face analyser implements the required messaging components and provides output in the arousal and valence dimensional space. The voice analyser has new components to enable software integration and is being refined through a spiral developmental process. A new release of the analyser is currently evaluated during recent focus group meetings and a new vocal expression evaluation system has been developed. The body gesture analyser can provide ActiveMQ messages and infrastructure to integrate the subsystem in the platform. the gesture analyser also refined the feature extraction modules and the emotion recognition system. Further, interactive training program lessons and expression games on emotions  by using the integrated 
modalities were developed. The content to be used by the general platform and the subsystems has been identified and created. The video material recorded as examples has been extensively validated such that out of 60 raters, at least half agree. Two surveys were initially developed and piloted to ensure the validation method chosen was methodologically sound. 
The evaluation protocols for the upcoming randomised control clinical trial are prepared. As an outlook, evaluation is needed, providing the basis for the further refinement of the analysers, and fine-tuned interplay with the platform.

\section*{Acknowledgement}
The research leading to these results has received funding from the European Community's Seventh Framework Programme (FP7/2007-2013) under grant agreement No.\ 289021 (ASC-Inclusion).\\

\section{Additional Authors}
Ofer Golan, Shimrit Fridenson, Shahar Tal (Department of Psicology, Bar-Ilan University, Ramat-Gan, Israel, email: \texttt{ofer.golan@biu.ac.il})
Shai Newman, Noga Meir, Roi Shillo (Compedia Ltd, Ramat-Gan, Israel, email: \texttt{newmans@compedia.net})
Antonio Camurri, Stefano Piana, Alessandra Staglian{\`o} (InfoMus Lab, DIST, University of Genoa, Genoa, Italy, email: \texttt{antonio.camurri@unige.it})
Sven B{\"o}lte, Daniel Lundqvist, Steve Berggren (Center of Neurodelopmental Disorders, Karolinska Intituted, Stockholm, Sweden, email: \texttt{sven.bolte@ki.se})
Aurelie Baranger and Nikki Sullings (Autism-Europe aisbl, Brussels, Belgium, email: \texttt{aurelie.baranger@autismeurope.org}).

%
%
%
%
%
\balance


\normalsize

%

\bibliographystyle{acm-sigchi}
\let\oldbibliography\thebibliography
\renewcommand{\thebibliography}[1]{%
\oldbibliography{#1}%
\setlength{\itemsep}{-2pt}%
}
\bibliography{ASC}

\begin{thebibliography}{10}

\bibitem{attwood1998asperger}
Attwood, T.
\newblock {\em Asperger's syndrome: A guide for parents and professionals}.
\newblock Jessica Kingsley Pub, 1998.

\bibitem{3d_6247980}
Baltrusaitis, T., Robinson, P., and Morency, L.
\newblock 3d constrained local model for rigid and non-rigid facial tracking.
\newblock In {\em Computer Vision and Pattern Recognition (CVPR), 2012 IEEE
  Conference on} (2012), 2610--2617.

\bibitem{baron1997mindblindness}
Baron-Cohen, S.
\newblock {\em Mindblindness: An essay on autism and theory of mind}.
\newblock MIT press, 1997.

\bibitem{baron2009prevalence}
Baron-Cohen, S., Scott, F.~J., Allison, C., Williams, J., Bolton, P., Matthews,
  F.~E., and Brayne, C.
\newblock Prevalence of autism-spectrum conditions: Uk school-based population
  study.
\newblock {\em The British Journal of Psychiatry 194}, 6 (2009), 500--509.

\bibitem{baron2005empathizing}
Baron-Cohen, S., Wheelwright, S., Lawson, J., Griffin, R., Ashwin, C.,
  Billington, J., and Chakrabarti, B.
\newblock Empathizing and systemizing in autism spectrum conditions.
\newblock {\em Handbook of autism and pervasive developmental disorders 1\/}
  (2005), 628--639.

\bibitem{bauminger2005social}
Bauminger, N., Edelsztein, H.~S., and Morash, J.
\newblock Social information processing and emotional understanding in children
  with ld.
\newblock {\em Journal of learning disabilities 38}, 1 (2005), 45--61.

\bibitem{Beaumont2008}
Beaumont, R., and Sofronoff, K.
\newblock A multi-component social skills intervention for children with
  asperger syndrome: The junior detective training program.
\newblock {\em Journal of Child Psychology and Psychiatry 49}, 7 (2008),
  743--753.

\bibitem{bell1994dsm}
Bell, C.~C.
\newblock Dsm-iv: Diagnostic and statistical manual of mental disorders.
\newblock {\em JAMA: The Journal of the American Medical Association 272}, 10
  (1994), 828--829.

\bibitem{bernardopitz2001}
Bernard-Opitz, V., Sriram, N., and Nakhoda-Sapuan, S.
\newblock Enhancing social problem solving in children with autism and normal
  children through computer-assisted instruction.
\newblock {\em Journal of Autism and Developmental Disorders 31}, 4 (2001),
  377--384.

\bibitem{bishop2003}
Bishop, J.
\newblock The internet for educating individuals with social impairments.
\newblock {\em Journal of Computer Assisted Learning 19}, 4 (2003), 546--556.

\bibitem{bolte2006facial}
B{\"o}lte, S., Hubl, D., Feineis-Matthews, S., Prvulovic, D., Dierks, T., and
  Poustka, F.
\newblock Facial affect recognition training in autism: can we animate the
  fusiform gyrus?
\newblock {\em Behavioral neuroscience 120}, 1 (2006), 211.

\bibitem{boucher2003voice}
Boucher, J., Lewis, V., and Collis, G.~M.
\newblock Voice processing abilities in children with autism, children with
  specific language impairments, and young typically developing children.
\newblock {\em Journal of Child Psychology and Psychiatry 41}, 7 (2003),
  847--857.

\bibitem{brownlow2006constructing}
Brownlow, C., and O'Dell, L.
\newblock Constructing an autistic identity: As voices online.
\newblock {\em Journal Information 44}, 5 (2006).

\bibitem{celani1999understanding}
Celani, G., Battacchi, M.~W., and Arcidiacono, L.
\newblock The understanding of the emotional meaning of facial expressions in
  people with autism.
\newblock {\em Journal of autism and developmental disorders 29}, 1 (1999),
  57--66.

\bibitem{da2009emotion}
Da~Fonseca, D., Seguier, V., Santos, A., Poinso, F., and Deruelle, C.
\newblock Emotion understanding in children with adhd.
\newblock {\em Child Psychiatry \& Human Development 40}, 1 (2009), 111--121.

\bibitem{de2010conversational}
de~Marchena, A., and Eigsti, I.-M.
\newblock Conversational gestures in autism spectrum disorders: Asynchrony but
  not decreased frequency.
\newblock {\em Autism Research 3}, 6 (2010), 311--322.

\bibitem{deruelle2004spatial}
Deruelle, C., Rondan, C., Gepner, B., and Tardif, C.
\newblock Spatial frequency and face processing in children with autism and
  asperger syndrome.
\newblock {\em Journal of autism and developmental disorders 34}, 2 (2004),
  199--210.

\bibitem{Eyben13-RDI}
Eyben, F., Weninger, F., Gro{\ss}, F., and Schuller, B.
\newblock {Recent Developments in openSMILE, the Munich Open-Source Multimedia
  Feature Extractor}.
\newblock In {\em {Proceedings of the 21st ACM International Conference on
  Multimedia, MM 2013}}, ACM, ACM (Barcelona, Spain, October 2013), 835--838.

\bibitem{Eyben10-OTM}
Eyben, F., W\"ollmer, M., and Schuller, B.
\newblock {openSMILE -- The Munich Versatile and Fast Open-Source Audio Feature
  Extractor}.
\newblock In {\em {Proceedings of the 18th ACM International Conference on
  Multimedia, MM 2010}}, ACM, ACM (Florence, Italy, October 2010), 1459--1462.

\bibitem{golan2010enhancing}
Golan, O., Ashwin, E., Granader, Y., McClintock, S., Day, K., Leggett, V., and
  Baron-Cohen, S.
\newblock Enhancing emotion recognition in children with autism spectrum
  conditions: an intervention using animated vehicles with real emotional
  faces.
\newblock {\em Journal of autism and developmental disorders 40}, 3 (2010),
  269--279.

\bibitem{CambridgeJournals:420155}
Golan, O., and Baron-Cohen, S.
\newblock Systemizing empathy: Teaching adults with asperger syndrome or
  high-functioning autism to recognize complex emotions using interactive
  multimedia.
\newblock {\em Development and Psychopathology 18}, 02 (2006), 591--617.

\bibitem{golan2008reading}
Golan, O., Baron-Cohen, S., and Golan, Y.
\newblock The `reading the mind in films' task [child version]: Complex emotion
  and mental state recognition in children with and without autism spectrum
  conditions.
\newblock {\em Journal of Autism and Developmental Disorders 38}, 8 (2008),
  1534--1541.

\bibitem{golan2006cambridge}
Golan, O., Baron-Cohen, S., and Hill, J.
\newblock The cambridge mindreading (cam) face-voice battery: Testing complex
  emotion recognition in adults with and without asperger syndrome.
\newblock {\em Journal of autism and developmental disorders 36}, 2 (2006),
  169--183.

\bibitem{golan2007reading}
Golan, O., Baron-Cohen, S., Hill, J.~J., and Rutherford, M.
\newblock The `reading the mind in the voice' test - revised: a study of
  complex emotion recognition in adults with and without autism spectrum
  conditions.
\newblock {\em Journal of autism and developmental disorders 37}, 6 (2007),
  1096--1106.

\bibitem{golan2007assistive}
Golan, O., LaCava, P.~G., and Baron-Cohen, S.
\newblock Assistive technology as an aid in reducing social impairments in
  autism.
\newblock {\em Growing Up with Autism: Working with School-Age Children and
  Adolescents\/} (2007), 124--142.

\bibitem{grezes2009failure}
Gr{\`e}zes, J., Wicker, B., Berthoz, S., and De~Gelder, B.
\newblock A failure to grasp the affective meaning of actions in autism
  spectrum disorder subjects.
\newblock {\em Neuropsychologia 47}, 8 (2009), 1816--1825.

\bibitem{hobson1986autistic}
Hobson, R.~P.
\newblock The autistic child's appraisal of expressions of emotion.
\newblock {\em Journal of Child Psychology and Psychiatry 27}, 3 (1986),
  321--342.

\bibitem{jordan2010evolution}
Jordan, C.~J.
\newblock Evolution of autism support and understanding via the world wide web.
\newblock {\em Intellectual and developmental disabilities 48}, 3 (2010),
  220--227.

\bibitem{kaliouby2005real}
Kaliouby, R., and Robinson, P.
\newblock Real-time inference of complex mental states from facial expressions
  and head gestures.
\newblock {\em Real-time vision for human-computer interaction\/} (2005),
  181--200.

\bibitem{kasari2001social}
Kasari, C., Chamberlain, B., and Bauminger, N.
\newblock Social emotions and social relationships: Can children with autism
  compensate?
\newblock In {\em {Development and autism: Perspectives from theory and
  research}}, J.~Burack, T.~Charman, N.~Yirmiya, and P.~Zelazo, Eds. Lawrence
  Erlbaum Associates Publishers, Mahwah, NJ, 2001, 309--323.

\bibitem{kasari1990affective}
Kasari, C., Sigman, M., Mundy, P., and Yirmiya, N.
\newblock Affective sharing in the context of joint attention interactions of
  normal, autistic, and mentally retarded children.
\newblock {\em Journal of autism and developmental disorders 20}, 1 (1990),
  87--100.

\bibitem{kasari1993affective}
Kasari, C., Sigman, M., Yirmiya, N., and Mundy, P.
\newblock {\em Affective development and communication in young children with
  autism.}
\newblock Paul H. Brookes Publishing, 1993.

\bibitem{laban1971mastery}
Laban, R., and Ullmann, L.
\newblock {\em The mastery of movement.}
\newblock ERIC, 1971.

\bibitem{macdonald2006recognition}
Macdonald, H., Rutter, M., Howlin, P., Rios, P., Conteur, A.~L., Evered, C.,
  and Folstein, S.
\newblock Recognition and expression of emotional cues by autistic and normal
  adults.
\newblock {\em Journal of Child Psychology and Psychiatry 30}, 6 (2006),
  865--877.

\bibitem{Marchi12-SEA}
Marchi, E., Batliner, A., Schuller, B., Fridenzon, S., Tal, S., and Golan, O.
\newblock {Speech, Emotion, Age, Language, Task, and Typicality: Trying to
  Disentangle Performance and Feature Relevance}.
\newblock In {\em {Proceedings First International Workshop on Wide Spectrum
  Social Signal Processing (WS3P 2012), ASE/IEEE SocialCom 2012}}, ASE/IEEE,
  IEEE (Amsterdam, The Netherlands, September 2012).

\bibitem{Marchi12-EIT}
Marchi, E., Schuller, B., Batliner, A., Fridenzon, S., Tal, S., and Golan, O.
\newblock {Emotion in the Speech of Children with Autism Spectrum Conditions:
  Prosody and Everything Else}.
\newblock In {\em {Proceedings 3rd Workshop on Child, Computer and Interaction
  (WOCCI 2012), Satellite Event of INTERSPEECH 2012}}, ISCA, ISCA (Portland,
  OR, September 2012).

\bibitem{mccann2003prosody}
McCann, J., and Peppe, S.
\newblock Prosody in autism spectrum disorders: a critical review.
\newblock {\em International Journal of Language \& Communication Disorders
  38}, 4 (2003), 325--350.

\bibitem{moore2005collaborative}
Moore, D., Cheng, Y., McGrath, P., and Powell, N.~J.
\newblock Collaborative virtual environment technology for people with autism.
\newblock {\em Focus on Autism and Other Developmental Disabilities 20}, 4
  (2005), 231--243.

\bibitem{moore2000computer}
Moore, D., McGrath, P., and Thorpe, J.
\newblock Computer-aided learning for people with autism -- a framework for
  research and development.
\newblock {\em Innovations in Education and Training International 37}, 3
  (2000), 218--228.

\bibitem{world1993icd}
Organization, W.~H.
\newblock {\em The ICD-10 classification of mental and behavioural disorders:
  diagnostic criteria for research}.
\newblock World Health Organization, 1993.

\bibitem{parsons2000development}
Parsons, S., Beardon, L., Neale, H., Reynard, G., Eastgate, R., Wilson, J.,
  Cobb, S., Benford, S., Mitchell, P., and Hopkins, E.
\newblock Development of social skills amongst adults with asperger’s
  syndrome using virtual environments: the ‘as interactive’project.
\newblock In {\em Proc. The 3rd International Conference on Disability, Virtual
  Reality and Associated Technologies, ICDVRAT} (2000), 23--25.

\bibitem{parsons2002}
Parsons, S., and Mitchell, P.
\newblock The potential of virtual reality in social skills training for people
  with autistic spectrum disorders.
\newblock {\em Journal of Intellectual Disability Research 46}, 5 (2002),
  430--443.

\bibitem{paul2005perception}
Paul, R., Augustyn, A., Klin, A., and Volkmar, F.~R.
\newblock Perception and production of prosody by speakers with autism spectrum
  disorders.
\newblock {\em Journal of Autism and Developmental Disorders 35}, 2 (2005),
  205--220.

\bibitem{philip2010deficits}
Philip, R., Whalley, H., Stanfield, A., Sprengelmeyer, R., Santos, I., Young,
  A., Atkinson, A., Calder, A., Johnstone, E., Lawrie, S., et~al.
\newblock Deficits in facial, body movement and vocal emotional processing in
  autism spectrum disorders.
\newblock {\em Psychological medicine 40}, 11 (2010), 1919--1929.

\bibitem{pollak2000recognizing}
Pollak, S.~D., Cicchetti, D., Hornung, K., and Reed, A.
\newblock Recognizing emotion in faces: developmental effects of child abuse
  and neglect.
\newblock {\em Developmental psychology 36}, 5 (2000), 679.

\bibitem{schroeder2010emotion}
Schr{\"o}eder, M., Baggia, P., Burkhardt, F., Pelachaud, C., Peter, C., and
  Zovato, E.
\newblock Emotion markup language (emotionml) 1.0.
\newblock {\em W3C Working Draft 29\/} (2010), 3--22.

\bibitem{Schuller13-AIE}
Schuller, B., Marchi, E., Baron-Cohen, S., O'Reilly, H., Robinson, P., Davies,
  I., Golan, O., Friedenson, S., Tal, S., Newman, S., Meir, N., Shillo, R.,
  Camurri, A., Piana, S., B\"olte, S., Lundqvist, D., Berggren, S., Baranger,
  A., and Sullings, N.
\newblock {ASC-Inclusion: Interactive Emotion Games for Social Inclusion of
  Children with Autism Spectrum Conditions}.
\newblock In {\em {Proceedings 1st International Workshop on Intelligent
  Digital Games for Empowerment and Inclusion (IDGEI 2013) held in conjunction
  with the 8th Foundations of Digital Games 2013 (FDG)}}, B.~Schuller,
  L.~Paletta, and N.~Sabouret, Eds., ACM, SASDG (Chania, Greece, May 2013).

\bibitem{Schuller12-A2T}
Schuller, B., Valstar, M., Eyben, F., Cowie, R., and Pantic, M.
\newblock {AVEC 2012 -- The Continuous Audio/Visual Emotion Challenge}.
\newblock In {\em {Proceedings of the 14th ACM International Conference on
  Multimodal Interaction, ICMI}}, L.-P. Morency, D.~Bohus, H.~K. Aghajan,
  J.~Cassell, A.~Nijholt, and J.~Epps, Eds., ACM, ACM (Santa Monica, CA,
  October 2012), 449--456.

\bibitem{silver2001evaluation}
Silver, M., and Oakes, P.
\newblock Evaluation of a new computer intervention to teach people with autism
  or asperger syndrome to recognize and predict emotions in others.
\newblock {\em Autism 5}, 3 (2001), 299--316.

\bibitem{silverman2010speech}
Silverman, L.~B., Bennetto, L., Campana, E., and Tanenhaus, M.~K.
\newblock Speech-and-gesture integration in high functioning autism.
\newblock {\em Cognition 115}, 3 (2010), 380--393.

\bibitem{stevens2001recognition}
Stevens, D., Charman, T., and Blair, R.~J.
\newblock Recognition of emotion in facial expressions and vocal tones in
  children with psychopathic tendencies.
\newblock {\em The Journal of genetic psychology 162}, 2 (2001), 201--211.

\bibitem{tanaka2010}
Tanaka, J.~W., Wolf, J.~M., Klaiman, C., Koenig, K., Cockburn, J., Herlihy, L.,
  Brown, C., Stahl, S., Kaiser, M.~D., and Schultz, R.~T.
\newblock Using computerized games to teach face recognition skills to children
  with autism spectrum disorder: the let’s face it! program.
\newblock {\em Journal of Child Psychology and Psychiatry 51}, 8 (2010),
  944--952.

\end{thebibliography}
\end{document}